# Static Quantized Radix-2 FFT/IFFT Processor for Constraints Analysis

Rozita Teymourzadeh, Memtode Jim Abigo, Mok Vee Hoong

*Faculty of Engineering, Architecture & Built Environment, UCSI University, Kuala Lumpur, Malaysia*

rozita_teymourzadeh@yahoo.com, jimdprofit@gmail.com, jimmy@ucsi.edu.my , Jalan Choo Lip Kung, Taman Taynton View, 56000 Kuala Lumpur, WP Kuala Lumpur, Malaysia . Phone:+60391018880. Fax: +60391323663

# Static Quantized Radix-2 FFT/IFFT Processor for Constraints Analysis

This research work focuses on the design of a high-resolution fast Fourier transform (FFT) /inverse fast Fourier transform (IFFT) processors for constraints analysis purpose. Amongst the major setbacks associated with such high resolution, FFT processors are the high power consumption resulting from the structural complexity and computational inefficiency of floating-point calculations. As such, a parallel pipelined architecture was proposed to statically scale the resolution of the processor to suite adequate trade-off constraints. The quantization was applied to provide an approximation to address the finite word-length constraints of digital signal processing (DSP). An optimum operating mode was proposed, based on the signal-to-quantization-noise ratio (SQNR) as well as the statistical theory of quantization, to minimize the trade-off issues associated with selecting the most application-efficient floating-point processing capability in contrast to their resolution quality.

1.  Introduction

Discrete Fourier Transform (DFT) is amongst the most fundamental operations in digital signal processing. However, the widespread uses of DFTs make its computational requirements an important issue. The direct computation of the DFT requires approximately $N^2$ operations where N is the transform size. The breakthrough of Cooley-Tukey (CT) FFT comes from the fact that it reduces the complexity to an order of $N\log_2 N$ operations. The FFT is therefore an efficient algorithm to compute the DFT and its inverse (IDFT). It has several applications in the field of signal processing including the real-time processing of wireless time-domain and frequency-domain signals especially for use in Orthogonal Frequency Division Multiplexing (OFDM) systems such as Digital Video Broadcasting (DVB), Digital Subscriber Line (xDSL) and WiMAX (IEEE 802.16) [1-4]. These applications require large-point FFT processing, such as 1024/2048/8192-point, FFTs for multiple carrier modulation.

Many FFT algorithms based on the CT decomposition such as radix-$2^2$, radix-$2^3$, radix-4, radix-(4+2), prime-factor as well as split-radix algorithms, have been proposed using the complex mathematical relationship to reduce the hardware complexity. The computational complexity and the hardware requirements are greatly dependent on the algorithm in use [2]. The conventional parallel architecture poses issues related to hardware cost, complexity, power consumption and is not easily flexible to meet other design constraints. As such, different architectures to efficiently map the different FFT algorithms to hardware have been proposed [3].

A first approach for these implementations concerns time non-critical applications and has small hardware requirements, but it needs a significantly large number of clock cycles to compute a full FFT. For example, in [4] one butterfly unit is used for all computations and $N+N.\log_2 N$ clock cycles are required for the computation of the FFT. A second implementation approach is for speed demanding applications, where one butterfly unit is used for each decimation stage of a radix-2 FFT [5].

A pipeline architecture based on the constant geometry radix-2 FFT algorithm, which uses $\log_2 N$ complex-number multipliers (more precisely butterfly units) and is capable of computing a full N-point FFT in N/2 clock cycles has been proposed in 2009 [8]. However, this architecture requires a large amount of delay elements (memory size of $N.\log_2 N$ samples) and a quite complicated switching mechanism for the routing of the data. This commonly used pipeline architecture is characterized by continuous processing of input data as well as being highly regular, making it straight forward to automatically generate FFTs of various lengths [6]. All these developments have introduced their own disadvantages, in addition to the age-long finite word-length effects of digital circuitry [7-9]. This paper thus, uses the pipeline architecture [1],[2] to propose a model for the analysis of important design constraints like the finite word-length effects and amount of resolution needed to achieve the appropriate SNR [8-10] for the desired design needs using the statistical tools for the analysis of a range of feasible resolution.

## 2. Design Architecture

### A. Algorithm development of the decimation-in-time (DIT) radix-p FFT

The DFT of an N-point sequence x[n] is given by:

$$X[k] = \sum_{n=0}^{N-1} x[n] W_N^{kn} \quad \text{For } k = 0, 1, 2,\ldots,N-1 \quad (1)$$

Where $W_N = e^{-j\left(\frac{2\pi}{N}\right)}$.

Consider the general formula of the DIT radix-p FFT as follows:

$$X\left[k + r\left(\frac{N}{p}\right)\right] = \sum_{n=0}^{\frac{N}{p}-1}\left(\sum_{j=0}^{p-1} x[pn+j] \cdot W_p^{jr} W_N^{jk}\right) W_{\frac{N}{p}}^{nk} \quad (2)$$

for k = 0,1,2,…,N/p-1 and r = 0,1,2,…,p-1. Using the above decomposition, the DFT can be reduced successively to N/p p-point DFTs. In general, this process can be repeated m times and therefore there are totally m stages in the implementation of the DFT.

### B. Parallel Architecture

The computational structure of a butterfly unit is shown in Figure 1. It is the fundamental computational of the parallel architecture.

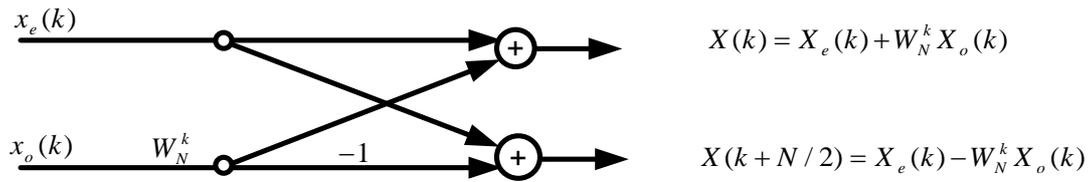

$$X(k) = X_e(k) + W_N^k X_o(k)$$

$$X(k+N/2) = X_e(k) - W_N^k X_o(k)$$

Figure 1. Radix-2 butterfly unit

The butterfly unit requires a complex multiply and two complex additions. Therefore, it takes a total of (N/2) Log$_2$N complex multiplies and Nlog$_2$N complex additions to compute all N-point DFT samples. An 8-point Radix-2 DIT FFT requires N/2 butterfly units per stage for all m stages. The input bit-reversed and inter-stage index routing gets even more complicated as the size of the unit N increases.

For larger butterflies (N > 26), the processor becomes extremely complex and slow. Hence, a simpler and faster architecture is then required. Therefore, the proposed system was designed and simulated to reduce the system complexity by using control signals. Figure 2, shows the overall pipelined system structure and its designed control signals.

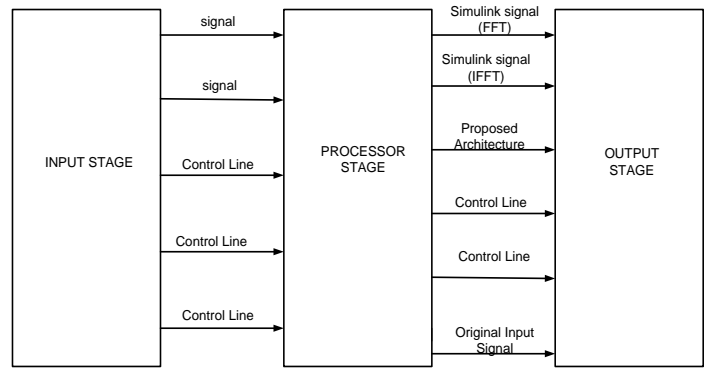

Figure 2. Proposed over view of pipelined system algorithm

## C. *Pipeline Architecture*

The same butterfly unit can perform the N/2 butterfly operations computed in every stage sequentially. Since the two inputs of a next butterfly unit of a stage are provided from the output of the butterfly unit of the previous stage at different time points, a shuffling unit is inserted between two successive butterfly units in order to route these outputs to the corresponding inputs of the next stage. Figure 3 shows the inner layer of proposed FFT processor design.

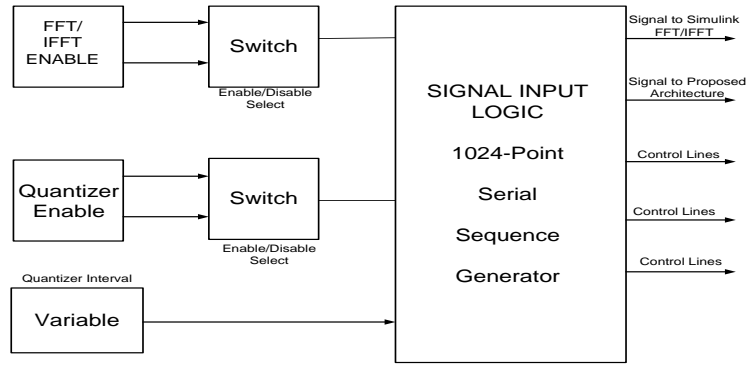

Figure 3. Proposed inner layer of FFT/IFFT Processor

The signal input is inserted at the control signal to program the processor functionality. The control signals are to select FFT or IFFT calculation, while the other enables and disables the quantization of the twiddle factors. Figure 4 illustrates the 10 stages butterfly for 1024-point pipeline FFT/IFFT processor.

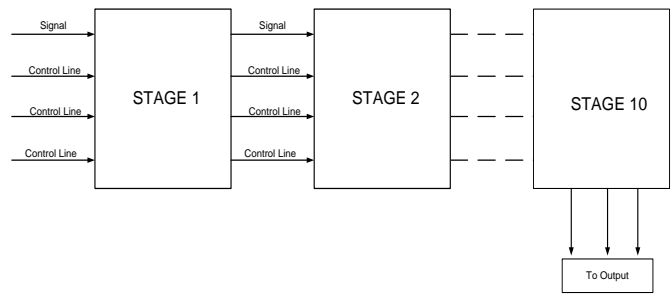

Figure 4. Ten-Stage 1024-point pipeline processor

In the proposed design, excluding the input stage, the rest of stages consist of the twiddle factors, the shuffling unit and a floating-point quantize model. The interval of the quantize unit for each stage is preset statically and this is used to vary the bit-resolution of the processor. Figure 5 shows the flowchart of overall system operation while quantizing is applied.

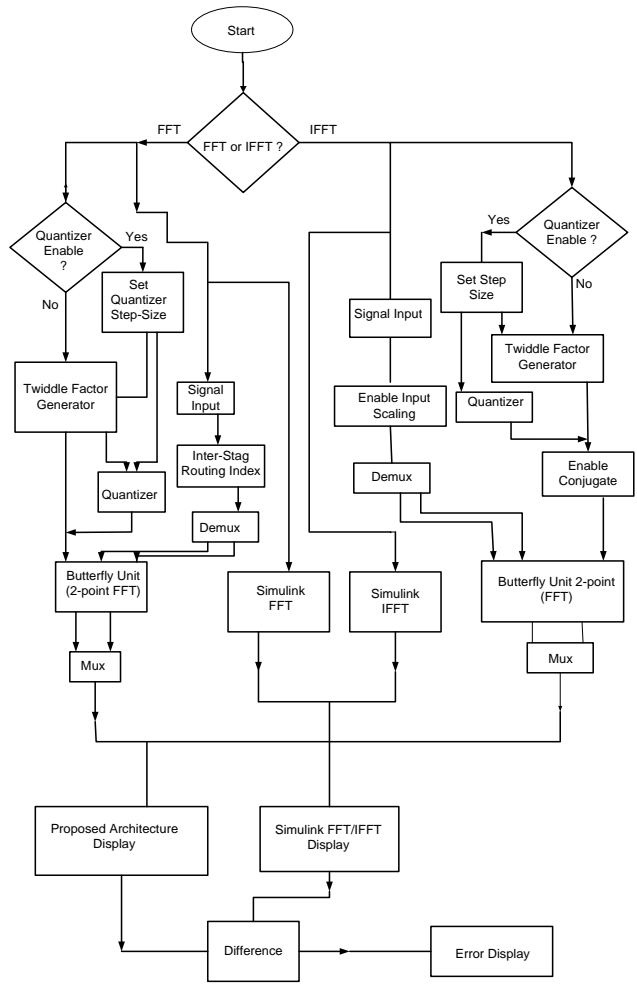

Figure 5. Proposed overall system operation

The IFFT computation uses the same fundamental Radix-2 DIT Butterfly unit. However, the input is scaled by the factor of N (1024). These discrete input values are then sent through the processor stage, which performs the same operation except that the conjugate of the twiddle factors are used instead. The output stage simply compares the results of the proposed FFT/IFFT Processor with the idle FFT/IFFT processor and their difference is observed as system error that will be analysed in the next chapter.

## 3. Statistical Theory of Quantization

### A. *Uniform Quantization*

One would expect that quantization has a similar effect on functions of the amplitude as sampling has on functions of time. Quantization is an operation on signals that is represented as a "staircase" function. The probability of each discrete output level equals the probability of the input signal occurring within the associated quantum band. For example, the probability that the output signal has the value zero equals the probability that the input signal falls between $\pm q/2$, where q is the quantization box size [8]. Figure 6 shows the model of quantizing

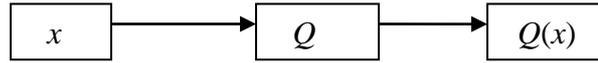

Figure 6. Uniform Quantize Model

The quantize error (*h*) is given as

$$h = x - Q(x) \qquad (3)$$

If x and h are real, with probability density function (PDF) as Px(.), then the quantization error variance is

$$\sigma_h^2 = E\{h^2\} = \int_{-\infty}^{\infty} h^2 P_h(h)\,dh = \sum_{k=1}^{L}\int_{x_k}^{x_{k+1}} (x - Q(x))^2 p_x(x)\,dx \qquad (4)$$

$$\sigma_h^2 = \int_{-\frac{q}{2}}^{\frac{q}{2}} h^2 \frac{1}{q}\,dq = \frac{q^2}{12} = \frac{1}{3} x_{max}^2\, 2^{-2b} \qquad (5)$$

$$SNR(dB) = 10\log_{10}\left(\frac{\sigma_x^2}{\sigma_h^2}\right) \qquad (6)$$

where *q* is the quantization interval, b is the number of bits. Quantization noise is defined as the difference between the output and input of the quantizer.

Since the quantized unit is designed in the proposed processor to enhance the calculations, Figure 7 illustrates a plot of the error versus the number of bits for the uniform quantizer, while Figure 8 shows a comparison of the mean, standard deviation as well as variance of the uniform quantization model achieved by the proposed FFT Processor.

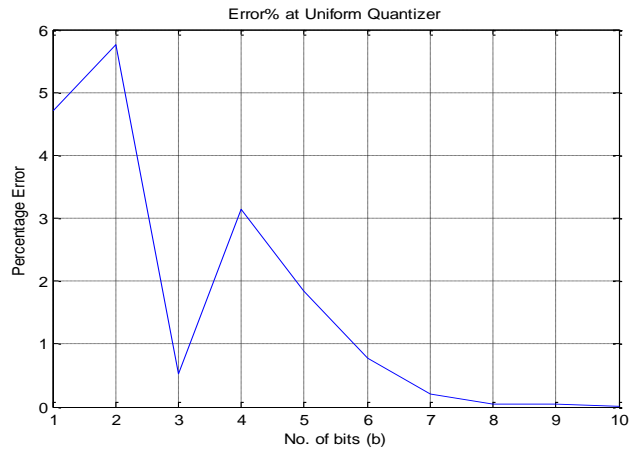

Figure 7. Error of the Uniform Quantization

The probability of getting a given error value is the sum of probabilities of all the quantization boxes. The uniform quantize model performs uniform quantization on the signal input and thus, a linear signal is required at the input. Fixed-point numbers are considered linear since the Radix point remains fixed. However, this research was focused on floating-point numbers hence; the response of the uniform quantize to floating point input was observed. As result, the quantization noise was increased. Equation (6) gives an expression for the SNR of the quantizer using the ratio of the variances of the input to noise.

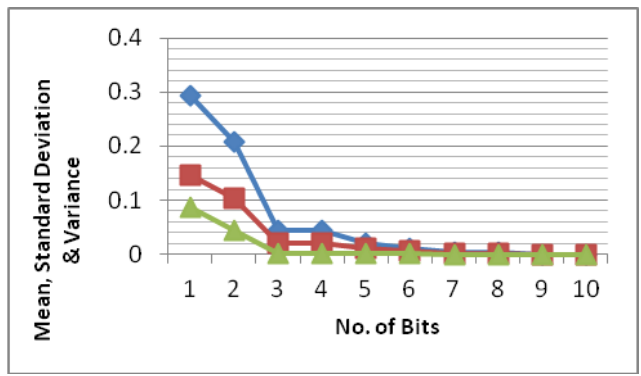

Figure 8. Measured dispersion of uniform quantizer

## B. Non-Uniform Quantization

The uniform quantization model is usually not used for floating-point quantization due to the overall non-uniform characteristic of the latter. Quantization of floating-point numbers is carried out only on the mantissa hence; it is more relevant to consider the relative error ε caused by the quantization process. The relative error defined in terms of the numerical values of the quantized floating-point number $Q(x) = 2^e Q(M)$ and the un-quantized number $x = 2^e M$ is given as

$$\varepsilon = \frac{(Q(x)-x)}{x} = \frac{(Q(M)-M)}{M} = \frac{\alpha}{M} \qquad (7)$$

It is possible however, to represent the floating-point quantizer using a combination of a compressor, a uniform quantizer and an expander. Figure 9 shows the non-uniform quantized model.

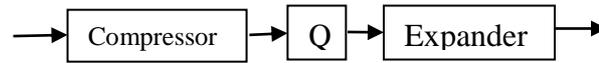

Figure 9. Non-uniform quantization model

$$\sigma_\varepsilon^2 = E\{\varepsilon^2\} = \frac{2}{q} \int_{1/2}^{1} \int_{-q/2}^{q/2} \frac{\alpha^2}{M^2 \, d\alpha} dM \qquad (8)$$

An expression for the variance is shown in equation (8). The variance from the floating-point quantization equals half that obtained from the uniform quantization which is a generally preferred characteristic. Figure 10 determines that the stability of the processor performance such that no variation occurs when the number of bits is 7 bit, unlike that of the uniform quantization which attains this stability at bit position eight 8, as shown in Figure 7. That is the advantage of the system while modelling the floating-point structure.

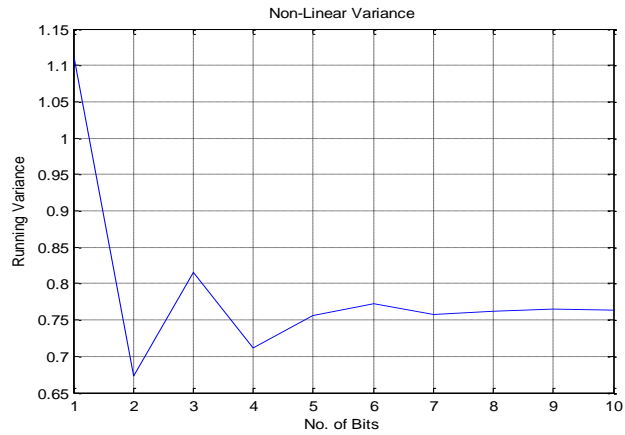

Figure 10. Error variance of the non-uniform quantized model

In addition, bit position 2 of Figure 10 gives minimum swing before stability, contrary to that of Figure 7, which occurs at bit position 3. This minimum swing gives a false minimum error position and is used for less sensitive applications in which minimum error is not important. Figure 11 illustrates the comparison between mean, standard deviation when number of bit increased.

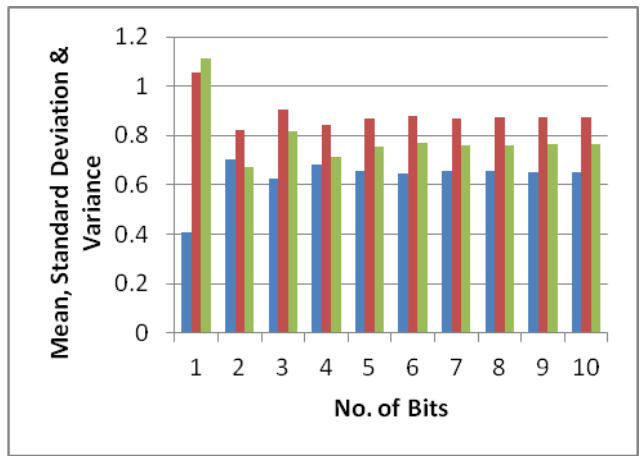

Figure 11. Comparison of the measured dispersion of non-uniform quantization

## 4. Discussion

Statistical parameters like the mean, standard deviation and variance were used to analytically develop expressions for the variation in error as a function of the quantization interval. These parameters were also known to have relationships with the SQNR. The percentage error of the non-uniform quantization generally decreased with an increase in the quantizer interval. As such, the SQNR is increased with respect to the quantization step size. The same general trend was observed in the uniform quantization, as well as the FFT and IFFT results. Figure 12 shows the error variation when the bit resolution increased.

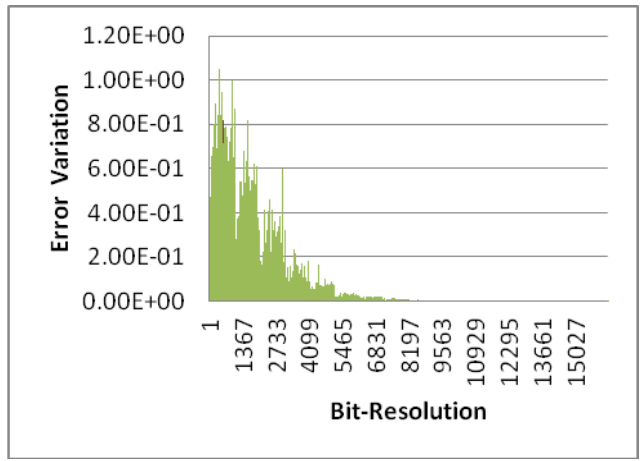

Figure 12. Error variation of the FFT processor

The general trend observed from the results indicates that the measured dispersion can only be valuable when they are used alongside the mean since the mean actually provides the benchmark for understanding the decreasing trend. Figure 13 shows the mean standard deviation and variance when the number of bit increased.

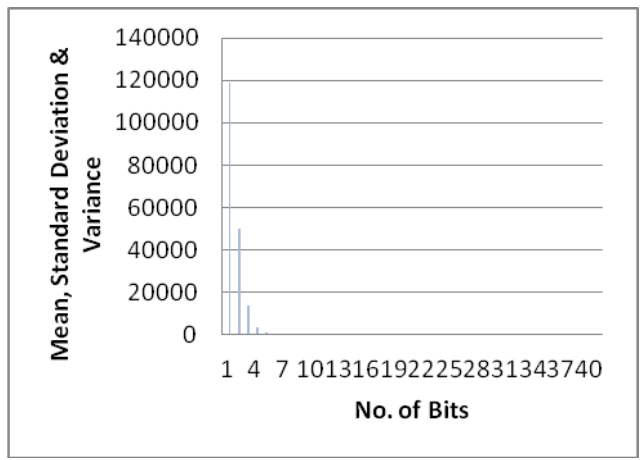

Figure 13. Measured dispersion of FFT/IFFT processor

Therefore, the variance is decreased as quantization interval increased. Hence, the variance is inversely proportional to the percentage error, and as such, inversely proportional to the SQNR. This provides experimental proof to the theoretical models given earlier and provides a benchmark for the trade-off between the SQNR and the resolution. In other words, bit number increasing has significant effect on resolution improvement. As shown in Figure 14 the error variation also decreased as bit resolution increased. In addition, it proves that higher number of data point in FFT processor will leads to have high accuracy spectrum measurement.

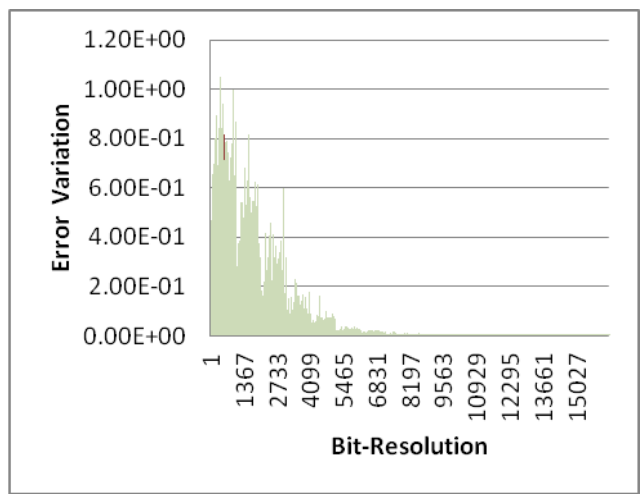

Figure 14. Error variation of IFFT processor

## 5. Conclusion

The percentage error and all the measured dispersion were found to decrease as the bit-resolution increased. This shows how the SQNR improves with bit-resolution. Although the power requirement for such SQNR systems are high, the proposed architecture provides an ease in the trade-off decision between the SQNR, power requirement and bit-resolution of the Radix-2 FFT/IFFT processor.

**References**

[1] Meletis, C. Bougas, P. Economakos, G. Kalivas, Kaimal Pekmestzi, P. (2005), 'High Speed Pipeline Implementation of Radix-2 DIF Algorithm', *Proceedings of World Academy of Science, Engineering and Technology*. Vol. 2, ISSN 1307-6884, pp. 1-4.

[2] Curtis, T. Curtis, M. (2004), 'High Performance Digital Signal Processing, *IOA Conference on Sonar Signal Processing'*, *Lough borough*, pp. 1-10.

[3] Mohamed Ismail, M. Rangachar, M.J.S. Paradesi, Ch. D. Rao, V. (2010), 'VLSI Implementation and Performance Analysis of Efficient Mixed-Radix 8-2 FFT Algorithm with Bit Reversal for the Output Sequences'. *Journal of Theoretical and Applied Information Technology*, pp. 164-168.

[4] Luis Diaz de Cerio, Miguel Valero-Garcia and Antonio Gonzalez. (1985), 'Efficient FFT on Torus Multicomputers: A Performance Study'. Universitat Politecnica de Catalunya, Barcelona (Spain), pp. 1 -10.

[5] Chih-Peng and Guo-An Su. (2006), 'A Grouped Fast Fourier Transform Algorithm Design for Selective Transformed Outputs'. Depart of Electrical, National Chung Hsing University, Taiwan, pp 1 – 8.

[6] Istvan Kollar, (1994), 'Bias of Mean Value and Mean Square Value Measurements Based on Quantized Data', *IEEE Transactions on Instrumentation and Measurement*, Vol 43, No 5, University of Stanford University, U.S.A, pp 733 – 739.

[7] Jonathan Greene, Robert Cooper. (2005), 'A Parallel 64K Complex FFT Algorithm for the IBM/Son/Toshiba Cell Broadband Engine Processor', *Mercury Computer Systems, Chelmsford MA*, pp 1 – 7.

[8] Ahmed Saeed, Elbably, M. Abdelfadeel, G. and Eladawy, M.I. (2009), 'Efficient FPGA Implementation of FFT/IFFT Processor', *International Journal of Circuits, Systems and Signal Processing*, *Chelmsford MA*, Vol. 3(3): 103-110.

[9] Satorius, E.H. Grimm, M.J. Zimmerman, G.A. and Wilck, H.C. (1986), 'Finite Word length Implementation of a Megachannel Digital Spectrum Analyzer', *TDA Progress Report*, pp 42 –86.

[10] Pel-Yun Tsai, Chia-Wei Chen and Meng-Yuan Huang (2011), 'Automatic IP Generation of FFT/IFFT Processors with Word-Length Optimization for MIMO-OFDM Systems', National Central University, Taiwan, pp 1 – 15.

[11] Wei-Hsin Chang and Truong Nguyen (2007), ' Integer FFT Optimized Coefficient Sets', Department of Electrical and Computer Engineering, UCSD, La Jolla, California, pp 1 – 4.

[12] Cheng-Yeh Wang, Chih-Bin Kuo, and Jing-Yang Jou. (2007), 'Hybrid Word-Length Optimization Methods of Pipelined FFT Processors', *IEEE Transactions on Computers*, Vol 56(8):1 – 14.

[13] Shannon, C. E. (1949), 'Communication in the presence of noise', *Proc. IRE*,Vol. 47, pp. 10-21.

[14] Sripad, B. and Snyder, D. L. A. (1917), ' Necessary and sufficient condition for quantization errors to be uniform and white', *IEEE Trans. Acoust, Speech, Signal Processing,* Vol. ASSP- 25(5):442-448.

[15] Kontro, J. Kalliojiirvi, K. and Neuvo, Y. (1992), 'Floating-point arithmetic in signal processing'. *In Proc. IEEE Int. Symp. Circuits and Systems*, *San Diego, CA,* , 92CH3139-3, vol. 4, pp. 1784-1791.

[16] The Institute of Electrical and Electronics Engineers IEEE Standard, (1987), 'IEEE Standard for binary floating-point nithmetic', *ANSVIEEE Standard 754-1985,* New York.

[17] Widrow, B. (1956), 'A study of rough amplitude quantization by means of Nyquist sampling theory' Sc.D. thesis, Department of Electrical Engineering, MIT.